\documentstyle[prl,aps,floats,epsf,color]{revtex}
\begin{document}
\baselineskip=12pt
\def\be{\begin{equation}}
\def\ee{\end{equation}}
\def\bea{\begin{eqnarray}}
\def\eea{\end{eqnarray}}
\def\E{{\rm e}}
\def\bearst{\begin{eqnarray*}}
\def\eearst{\end{eqnarray*}}
\def\peleven{\parbox{11cm}}
\def\peffec{\peight{\bearst\eearst}\hfill\peleven}
\def\pspace{\peight{\bearst\eearst}\hfill}
\def\ptwelve{\parbox{12cm}}
\def\peight{\parbox{8mm}}
\twocolumn
[\hsize\textwidth\columnwidth\hsize\csname@twocolumnfalse\endcsname

\title
{Characteristic Angular Scales in Cosmic Microwave Background
Radiation}

\author{ F. Ghasemi$^1$, A. Bahraminasab$^2$, M. Sadegh
Movahed$^{3,4} $, Sohrab Rahvar$^{3,5}$,\\ K. R. Sreenivasan$^{6}$
M. Reza Rahimi Tabar$^{3,7}$}
\address{$^{1}$The Max Planck Institute for the Physics of Complex
Systems, Nöthnitzer Strasse 38, 01187 Dresden, Germany  }
\address{$^{2}$ Department of Physics, Lancaster University,
Lancaster LA1 4YB, United Kingdom }
\address{$^{3}$ Dep. of Physics, Sharif University of Technology, P.O. Box
11365-9161, Tehran, Iran}
\address{$^{4}$ Institute for Studies in theoretical Physics and
Mathematics, P.O.Box 19395-5531,Tehran, Iran}
\address{$^{5}$ Research Institute for Astronomy \& Astrophysics of
Maragha, P.O.Box 55134-441, Maragha, Iran}
\address{$^{6}$ ICTP, Strada Costiera 11, I-34100 Trieste, Italy}
\address{$^{7}$CNRS UMR 6202, Observatoire de la C$\hat o$te d'Azur, BP
4229, 06304 Nice Cedex 4, France} \vskip 1cm
 \maketitle
\begin{abstract}
We investigate the stochasticity in temperature fluctuations in the
cosmic microwave background (CMB) radiation data from {\it Wilkinson
Microwave Anisotropy Probe}. We show that the angular fluctuations
of the temperature is a Markov process with a {\it Markov angular
scale}, $\Theta_{\rm Markov}=1.01^{+0.09}_{-0.07}$. We characterize
the complexity of the CMB fluctuations by means of a Fokker-Planck
or Langevin equation and measure the associated Kramers-Moyal
coefficients for the fluctuating temperature field $T(\hat n )$ and
its increment, $\Delta T =T(\hat n_1) - T(\hat n_2)$. Through this
method we show that temperature fluctuations in the CMB has fat
tails compared to a Gaussian distribution.
\end{abstract}
\hspace{.3in}
\newpage
 ]
\section{Introduction}
The {\it Wilkinson Microwave Anisotropy Probe} (WMAP) mission is one
of the main experiments in the cosmic microwave background (CMB)
that determines the power spectrum of the temperature fluctuations
in the CMB with high accuracy. It has been shown that the universe
is geometrically flat, and that the dominant content of the universe
is an exotic dark energy which causes the expansion of the universe
to be accelerated \cite{af04,ben03,spe03,teg04}. The statistical
properties of the CMB radiation is an important tool for identifying
the appropriate cosmological model and determining the parameters of
the standard model \cite{pei03}. The traditional way of studying the
CMB data is through analyzing their angular power spectrum and
computing the two-point correlation function. In order to gain full
statistical information from the temperature fluctuations on the
last scattering surface, we should use the $n$-point joint
probability density function (PDF) or the $n$-point correlation
function.

At the same time, Gaussianity of the CMB anisotropy is an
important question which is linked with the theoretical
predictions made by the inflationary cosmology
\cite{mu81,haw82,guth85}. With current developments in
experimental CMB physics, we are now in a position to analyze very
large data sets that provide information about large patches of
the sky, measured with very high resolution and sensitivity. This
means that we can precisely test whether the CMB is Gaussian.
Tantalizing evidence of non-Gaussianity has been emerging in the
WMAP sky maps, using a variety of methods. The statistical
properties of the primordial fluctuations generated by the
inflationary cosmology are closely related to the anisotropy of
the CMB radiation. Thus, measurement of the possible
non-Gaussianity of the CMB data is a direct test of the inflation
paradigm. If the CMB anisotropy is Gaussian, then the angular
power spectrum, or the two-point correlation function, fully
specifies its statistical properties.

In this paper we study the statistical properties of the CMB
anisotropy through the $n$-point joint PDF of temperature
fluctuations in the CMB data. We show that the fluctuations
constitute a Markovian process with a Markov angular scale of
$\Theta_{\rm Markov}=1.01^{+0.09}_{-0.07}$, or in the Legendre
space, $l_{\rm Markov}=178.22^{+0.27}_{-0.21}$, at $1\sigma$
confidence level. Using the Markov properties of the CMB data, a
master equation for the angular evolution of the probability
distribution function - the Fokker-Planck (FP) equation - is
obtained. To derive the relation between the standard power-spectral
analysis and the Markov properties of the CMB data, we consider the
joint probability distribution $P(T_2, \hat n_2; T_1, \hat n_1)$
that describes the probability of finding simultaneously $T_1$ in
the direction, $\hat{n}_1=(\theta_1,\phi_1)$, and $T_2$ in the
direction $\hat{n}_2=(\theta_2,\phi_2)$. We then evaluate the
corresponding Kramers-Moyal (KM) coefficients, and compute the first
and second KM coefficients (i.e., the drift and diffusion
coefficients in the FP equation). We show that the higher-order
coefficients in the KM expansion are very small and can be ignored.
In addition, using the FP equation for the PDF, we show that
temperature fluctuations on the last scattering surface are
consistent with a Gaussian distribution. We also use the same
analysis to investigate temperature fluctuations in the northern and
southern hemispheres. The analysis detects some cold and hot spots
in the southern and northern hemisphere, respectively.

The organization of this paper is as follows: In Section
\ref{correlation} we determine the 2-point joint PDF of
temperature fluctuations and obtain the correlation angular scale.
In Section \ref{M_P} we introduce the mathematical property of the
PDF for the Markovian process and obtain the Markov angular scale
for the WMAP data. We also develop the master equation in the form
of a FP equation for the CMB data. In Section \ref{gaussian} we
investigate the (non)-Gaussian nature of the temperature
fluctuations in the CMB. The relation between the Markov and the
angular power spectrum for this set of data is obtained in Section
\ref{KMC}. The conclusions are presented in Section \ref{conc}.
\begin{center}
\begin{figure}
\vspace{-1cm} \epsfxsize=7truecm\epsfbox{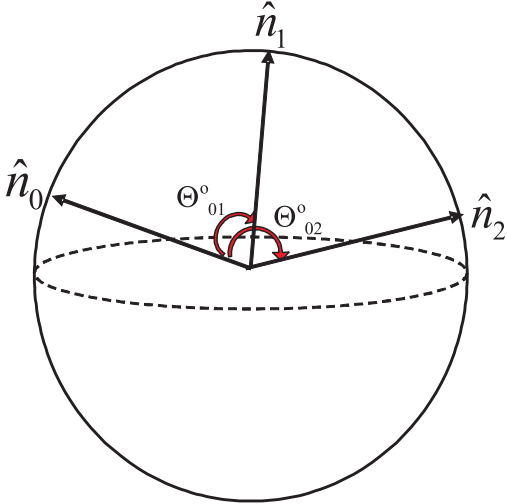} \narrowtext
\caption{ The arbitrary directions $\hat{n}_0$, $\hat{n}_1$ and
$\hat{n}_2$ in the sky to the last scattering surface.}
\label{sphere}
\end{figure}
\end{center}

\begin{center}
\begin{figure}
\vspace{1cm} \epsfxsize=9truecm\epsfbox{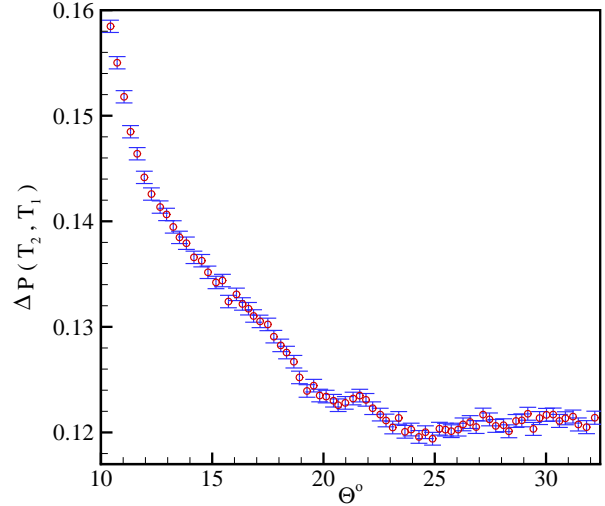} \narrowtext
\caption{$\Theta$ is the angular separation between two directions
$\hat n_1$ and $\hat n_2$. For $\Theta_{C}=24.86^o \pm 0.03$ we
obtain the minimum value for the $\Delta P(T_2,T_1)=|P( T_2,\hat
n_2;T_1,\hat n_1)-P(T_1,\hat n_1)P(T_2,\hat n_2)|$ vs $\Theta$.}
\label{f1}
\end{figure}
\end{center}

\section{Correlation angular scale by the joint PDF decomposition
analysis} \label{correlation} The WMAP instrument is composed of
10 differencing assemblies, spanning five frequencies from 23 to
94 GHz \cite{ben03b}. The two lowest frequency bands (K and Ka)
are primarily galactic foreground monitors, while the three
highest (Q, V, and W) are primarily cosmological bands. The full
width at half-maximum for the detectors is a function of the
frequency and ranges from $0.82^\circ$ at 23 GHz to $0.21^\circ$
at 94 GHz \cite{page03}. Here, we use the Internal Linear
Combination Map which is a weighted linear combination of the five
WMAP frequency maps. The weights are computed using the criterion
that minimize the galactic foreground contribution to the sky
signal. The resultant map provides a low-contamination image of
the CMB anisotropy.

Let the temperature fluctuation in the CMB data in the direction
of $\hat n =(\theta,\phi)$ be represented by ${\cal T}(\hat n)$,
and define $T(\hat{n}) \equiv ({\cal T}(\hat n) - \bar T)/\sigma$,
where $\bar T$ and $\sigma$ are the mean and variance of the
temperature fluctuations, respectively. For a direction $\hat n_0$
with $T(\hat n_0)=T_0$, at any $\hat n \neq \hat n_0$, $T(\hat n)$
is given by a PDF $P(T,\hat n)$ (see figure 1). This means that
$T(\hat n)$ for $\hat n \neq \hat n_0$ is a stochastic (and
possibly correlated) variable.

Complete information about the stochastic process is obtained from
the knowledge about all the possible $k$-point or, more precisely,
$k$-scale joint PDF (JPDF), $P(T_{k},\hat n_k;T_{k-1},\hat
n_{k-1};...;T_{1},\hat n_1)$, describing the probability of finding
simultaneously, $T_{1}$ in the direction $\hat{n}_{1}$, $T_{2}$ in
the direction $\hat{n}_{2}$, and so on up to $T_{k}$ on the
direction $\hat{n}_{k}$. Moreover, we can define the conditional
probability distribution function. For a given $T(\hat n_0)=T_0$,
the conditional probability of $T(\hat n)$ at the successive
positions $\hat n_1,\hat n_2, \cdots, \hat n_k$, to be $T(\hat n_1),
T(\hat n_2), \cdots, T(\hat n_k)$ is described by
\begin{eqnarray}
&&P_k ^1( T_k, \hat n_k; \cdots; T_1, \hat n_1 | T_0, \hat n_0 )
dT_k dT_{k-1} \cdots dT_1
=\nonumber\\
&&{\rm Prob} \{ T(\hat n_i) \in [ T_i, T_i + dT_i ] \hskip 0.2cm
\nonumber \\
&&{\rm for} \hskip 0.2cm i=1,2, \cdots k \hskip 0.2cm {\rm and}
\hskip 0.2cm T(\hat n_0) =T_0 \}.
\end{eqnarray}
In this notation, the superscript (for the $P$) represents the
number of conditions (e.g., here, we have one condition), while
the subscript denotes the number of stochastic variables in the
joint PDF. The $k$-point JPDF of the temperature fluctuations is
expressed by the product of the multiconditional PDFs as
\begin{eqnarray}\label{joint1}
&&P(T_{k-1},\hat n_{k-1};T_{k-2},\hat n_{k-2};...;T_{1},\hat
n_1;T_{0},\hat n_0)=\nonumber\\
&&P_1^{k-1}(T_{k-1},\hat n_{k-1}|T_{k-2},\hat
n_{k-2};...;T_{1},\hat n_1;T_{0},\hat n_0) \nonumber \\&& \times
P_1^{k-2}(T_{k-2},\hat n_{k-2}|T_{k-3},\hat n_{k-3};...;
T_{1},\hat n_1;T_{0},\hat n_0)\nonumber\\&&
\times...P_1^1(T_{1},\hat n_1|T_{0},\hat n_0)P(T_{0},\hat n_0)~.
\end{eqnarray}
where, $P_1^1(T_{1},\hat n_1|T_{0},\hat n_0)$ denotes a conditional
probability of finding temperature fluctuation $T_{1}$ in the
direction $\hat{n}_1$, under the condition that the temperature
$T_0$ in the direction $\hat{n}_0$ has been found. This conditional
probability can be expressed by 2-point JPDF, as: $ P_1^1(T_{1},\hat
n_1|T_{0},\hat
n_0)=\frac{P(T_1,\hat{n}_1;T_0,\hat{n}_0)}{P(T_0,\hat{n}_0)}$, this
result is important because, if the conditional PDF for the $T_1$ is
independent of $T_{0}$, then $P_1^1(T_{1},\hat n_1|T_{0},\hat
n_0)=P(T_{1},\hat{n}_1)$, in this case: $P_1^1(T_{1},\hat
n_1;T_{0},\hat n_0)=P(T_{1},\hat{n}_1)P(T_{0},\hat{n}_0)$.

To derive the first characteristic angular scales in the CMB, we
define the {\it correlation angular scale}, where the JPDF can be
decomposed into the production of two independent PDFs \cite{pee80}
as
\begin{equation}
P(T_2,\hat n_2; T_1, \hat n_1)|_{\Theta_{C}}=P(T_1,\hat n_1)
P(T_2,\hat n_2).
\end{equation}
Here, $\Theta$ is the angular separation between the two direction
$\hat n_1$ and $\hat n_2$, and we assumed the statistical isotropy
of the temperature fluctuations in the CMB in which the
correlation function depends only on the angle between the two
directions of $\hat n_1$ and $\hat n_2$ \cite{don05}. In Fig.
\ref{f1} we plot $\Delta P(T_2,T_1)=|P( T_2,\hat n_2; T_1, \hat
n_1) - P(T_1,\hat n_1) P( T_2,\hat n_2)|$ in terms of $\Theta$,
using the least square method,
\begin{eqnarray}
&&\chi^2=\int \frac{[P( T_2,\hat n_2; T_1, \hat n_1) - P(T_1,\hat
n_1) P( T_2,\hat n_2)]^2}{\sigma_{PDF}^2+\sigma_{2-{\rm joint}}^2}
dT_1dT_2,\nonumber
\end{eqnarray}
where the minimum value of $\chi^2$ corresponds to the angular
scale, at which the correlation disappears. Here, $\sigma_{PDF}^2$
and $\sigma_{2-{\rm joint}}^2$ are the variance of $P(T_1,\hat
n_1) P( T_2,\hat n_2)$ and $P( T_2,\hat n_2; T_1, \hat n_1)$,
respectively. Our analysis indicates that the minimum value of
$\Delta P(T_2,T_1)$ happens at the $\Theta_C=24.86^o \pm 0.03$
with $\chi^2_{\nu}=1.01$ ($\chi^2_{\nu}={\chi^2}/{\cal{N}}$, with
$\cal{N}$ being the number of degree of freedom).

\section{CMB data as a Markov Process: Least Squares Test}
\label{M_P} Now, let us check whether the CMB data follow a Markov
chain and, if so, measure the second characteristic angular scale
in the CMB, i.e., the Markov angular scale $\Theta_{\rm Markov}$ -
the scale over which the CMB data are Markov-correlated. In other
words, the Markov angular scale $\Theta_{\rm Markov}$ is the
minimum angular separation over which the data can be represented
by a Markov process \cite{fri98,jaf03,pei04}. The exact
mathematical definition of the Markov process is given
\cite{ris84} by
\begin{eqnarray}\label{mar1}
&&P (T_k,\hat n_k  | T_{k-1},\hat n_{k-1}; \cdots; T_1,\hat
n_1; T_0, \hat n_0) =\nonumber\\
&&\qquad \qquad P (T_k,\hat n_k | T_{k-1},\hat n_{k-1}).
\end{eqnarray}
Intuitively, the physical interpretation of a Markov process is that
it "forgets its past," or, in other words, only the most nearby
conditioning, say $(T_k,n_k)$, is relevant to the probability of
finding a temperature $T_k$ at $n_k$. In the Markov process the
ability to predict the value of $T(\hat n)$ will not be enhanced by
knowing its values in the steps prior to the most recent one.
Therefore, an important simplification that can be made for a Markov
process is that a conditional multivariate JPDF (Eq. \ref{joint1})
can be written in terms of the products of simple two-parameter
conditional PDFs \cite{ris84} as
\begin{eqnarray}
&&P (T_k,\hat n_k; T_{k-1},\hat n_{k-1}; \cdots; T_1,\hat
n_{1}| T_0, \hat n_{0}) =\nonumber\\
&& \qquad \qquad \prod_{i=1}^ k P(T_i,\hat n_{i} | T_{i-1}, \hat
n_{i-1}).
\end{eqnarray}

In what follows, we use the least-square method to determine the
Markov angular scale of the CMB temperature fluctuations. Testing
Eq. (\ref{mar1}) for large values of $k$ is out of our
computational capability; however, for $k=3$, where we have three
vectors pointing the 2D celestial sphere, the Markov condition is
as follows,
\begin{equation}\label{2p}
P(T_3,\hat n_3|T_2, \hat n_2; T_1, \hat n_1)=P(T_3,\hat n_3|T_2,
\hat n_2),
\end{equation}
where $\Theta_{ij}=\arccos(\hat n_i.\hat n_j)$ is the separation
angle of $\hat n_i$ and $\hat n_j$ and $\Theta_{31}>\Theta_{32},
\Theta_{21}$ (see Fig. \ref{sphere}). For simplicity, we let
$\Theta_{21}=\Theta_{32}$. A process is Markovian if Eq. (\ref{2p})
is satisfied for a certain angular separation which, in our
notation, is $\Theta_{32}$. We refer to it as the Markov angular
scale $\Theta_{\rm Markov}$).

In order to determine the Markov angular scale, we compare the
three-point PDF with that obtained based on the Markov process.
The three-point PDF, in terms of conditional probability
functions, is given by
\begin{eqnarray}\label{3p}
&&P(T_3,\hat n_3;T_2, \hat n_2; T_1, \hat n_1)=\nonumber\\
&&\qquad P(T_3,\hat n_3|T_2, \hat n_2;T_1, \hat n_1)P(T_2,\hat
n_2;T_1, \hat n_1).
\end{eqnarray}
Using the properties of the Markov process and substituting Eq.
(\ref{2p}) we obtain:
\begin{eqnarray}\label{2pm}
&&P_{Mar}(T_3,\hat n_3;T_2, \hat n_2; T_1, \hat
n_1)=\nonumber\\
&&\qquad P(T_3,\hat n_3|T_2, \hat n_2)P(T_2,\hat n_2;T_1, \hat
n_1).
\end{eqnarray}

\begin{figure}[t]
\vspace{-1cm} \epsfxsize=9truecm\epsfbox{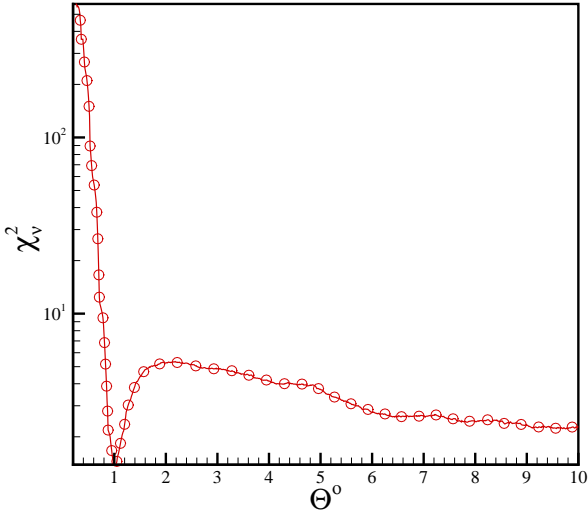} \narrowtext
\caption{$\chi^2_{\nu}$ in terms of angular scale $\Theta$. The
minimum value of $\chi^2_{\nu}=1.388$ corresponds to $\Theta_{\rm
Markov}=1.01^{+0.09}_{-0.07}$, with $1\sigma$ confidence level.}
\label{2}
\end{figure}

\begin{figure}[t]
\vspace{-1cm} \epsfxsize=9truecm\epsfbox{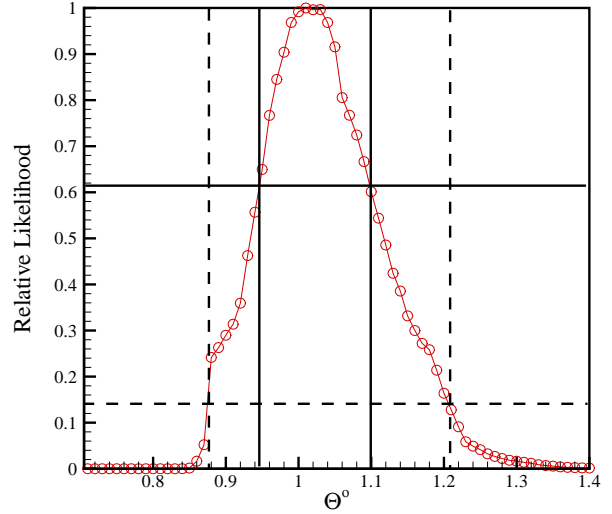} \narrowtext
\caption{Relative likelihood function of the Markov angular scale of
the CMB fluctuations, as a function of  $\Theta^{\circ}$. The
intersections of the curve with the horizontal solid and dashed
lines give the bounds with $1\sigma$ and $2\sigma$ confidence
levels, respectively. The minimum value of $\chi^2_{\nu}=1.38$
corresponds to $\Theta_{\rm Markov}=1.01^{+0.09}_{-0.07}$, with
$1\sigma$ and $\Theta_{\rm Markov}=1.01^{+0.19}_{-0.13}$, with
$2\sigma$ confidence level.} \label{3}
\end{figure}

In order to check the condition for the data being a Markov
process, we must compute the three-point JPDF through Eq.
(\ref{3p}) and compare the result with Eq. (\ref{2pm}). The first
step in this direction is to determine the quality of the fit
through the least-squared fitting quantity $\chi^2$ defined by
\begin{eqnarray}\label{chi}
&&\chi^2=\int dT_3 dT_2 dT_1[P(T_3,\hat n_3;T_2, \hat n_2; T_1,
\hat n_1)\nonumber\\&&-P_{\rm Mar}(T_3,\hat n_3;T_2, \hat n_2;
T_1, \hat n_1)]^2/ \left[\sigma_{3-{\rm joint}}^2+\sigma_{\rm
Mar}^2\right]
\end{eqnarray}
where $\sigma^2_{3-{\rm joint}}$ and $\sigma^2_{\rm Mar}$ are the
variances of $P(T_3,\hat n_3;T_2, \hat n_2; T_1, \hat n_1)$ and
$P_{\rm Mar}(T_3,\hat n_3;T_2, \hat n_2; T_1, \hat n_1)$,
respectively. To compute the Markov angular scale, we use the
Likelihood statistical analysis \cite{co04}. In the absence of a
prior constraint, the probability of the set of three-points JPDF
is given by a product of Gaussian functions:
\begin{eqnarray}
&&p(\arccos(\hat n_3,\hat
n_2))=\prod_{T_3,T_2,T_1}\frac{1}{\sqrt{2\pi(\sigma_{3-{\rm
joint}}^2
+\sigma_{\rm Mar}^2)}}\nonumber\\
&&\exp[-\frac{[P(T_3,\hat n_3;T_2, \hat n_2; T_1, \hat n_1)
-P_{\rm Mar}(T_3,\hat n_3;T_2, \hat n_2; T_1, \hat
n_1)]^2}{2(\sigma_{3-{\rm joint}}^2+\sigma_{\rm Mar}^2)}]\nonumber\\
\end{eqnarray}
This probability distribution must be normalized. Evidently, when,
for a set of values of the parameters, the $\chi^2$ is minimum the
probability is maximum. Figure \ref{2} shows the normalized
$\chi^2_{\nu}$ as a function of the angular length scale
$\Theta=\Theta_{32}\equiv \arccos(\hat n_3.\hat n_2)$. The minimum
value of $\chi^2_{\nu}$ is $1.38$, corresponding to $\Theta_{\rm
Markov}=1.01^{+0.09}_{-0.07}$ with $1\sigma$ confidence level.
Figure \ref{3} shows the likelihood function of Markov angular scale
of CMB.

The Markov nature of the CMB enables us to derive a master
equation - a FP equation - for the evolution of the PDF $P(T,\hat
n)$, in terms of, for example, the direction $\hat n$. One writes
Eq. (8) as an integral equation, which is well-known as the
Chapman-Kolmogorov (CK) equation:
\begin{equation}\label{ck}
P(T_3,\hat n_3| T_1, \hat n_1)=\int dT_2\;P(T_3,\hat n_3| T_2,\hat
n_2)\;P (T_2,\hat n_2| T_1,\hat n_1)\; \label{ck}
\end{equation}

\begin{figure}[t]
\vspace{-1cm} \epsfxsize=9truecm\epsfbox{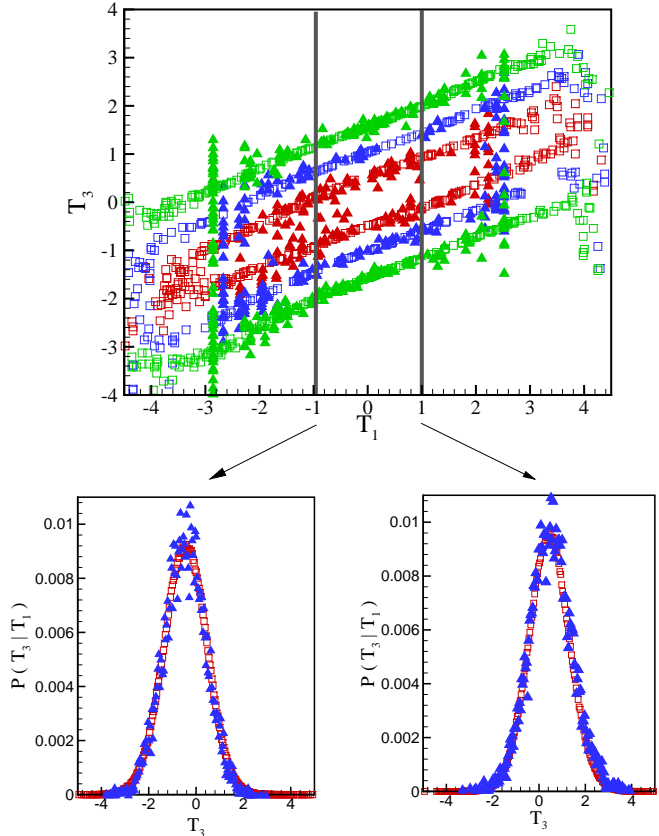} \narrowtext
\caption{Verification of the validity of the Chapman-Kolmogorov
equation (Eq. \ref{ck}) for the angular separation of $\hat n_1$
and $\hat n_2 $ being equal to the Markov angular scale. In the
upper panel we show the identification of left (filled symbol) and
right (open symbol) sides of Eq. (\ref{ck}) for three levels,
0.008 (red symbol), 0.005 (blue symbol), 0.002 (green symbol). The
conditional PDFs $P(T_3|T_1)$, for $T_1 =\pm\sigma$ are shown at
the lower panel. All the scales are measured in unit of the
standard deviation of the temperature fluctuations.} \label{f3}
\end{figure}

We checked the validity of the CK equation for describing the
angular separation of $\hat n_1$ and $\hat n_2 $ being equal to the
Markov angular scale. This is shown in Fig. \ref{f3}. In the upper
panel we show the identification of the left (filled symbol) and
right (open symbol) sides of Eq. (\ref{ck}) for three levels, 0.008
(red symbol), 0.005 (blue symbol), 0.002 (green symbol). The
conditional PDF $P(T_3|T_1)$, for $T_1=\pm\sigma$ are shown in the
lower panel. All the scales are measured in unit of the standard
deviation of the temperature fluctuations. The CK equation,
formulated in differential form, yields the following Kramers-Moyal
(KM) expansion \cite{ris84},
\begin{equation}\label{fokker1}
\label{f22} \frac{\partial}{\partial\phi}
P(T,\phi)=\sum_{n=1}^{\infty} (- \frac{\partial}{\partial T})^n
[D^{(n)}(T,\phi) P(T,\phi)],
\end{equation}
where $D^{(n)}(T,\phi)$ are called as the KM coefficients. These
coefficients can be estimated directly from the moments
$(M^{(n)})$ and the conditional probability distributions as:
\begin{eqnarray}\label{km}
&& D^{(n)}(T,\phi)=\frac{1}{n!}\hskip .2cm \lim_{\Delta \phi \to
0}M^{(n)},
\cr\nonumber\\
&& M^{(n)}=\frac{1}{\Delta \phi}\int dT'(T'-T)^n P(T',\phi+\Delta
\phi|T, \phi). \label{d12}
\end{eqnarray}

\begin{figure}[t]
\vspace{-1cm} \epsfxsize=9truecm\epsfbox{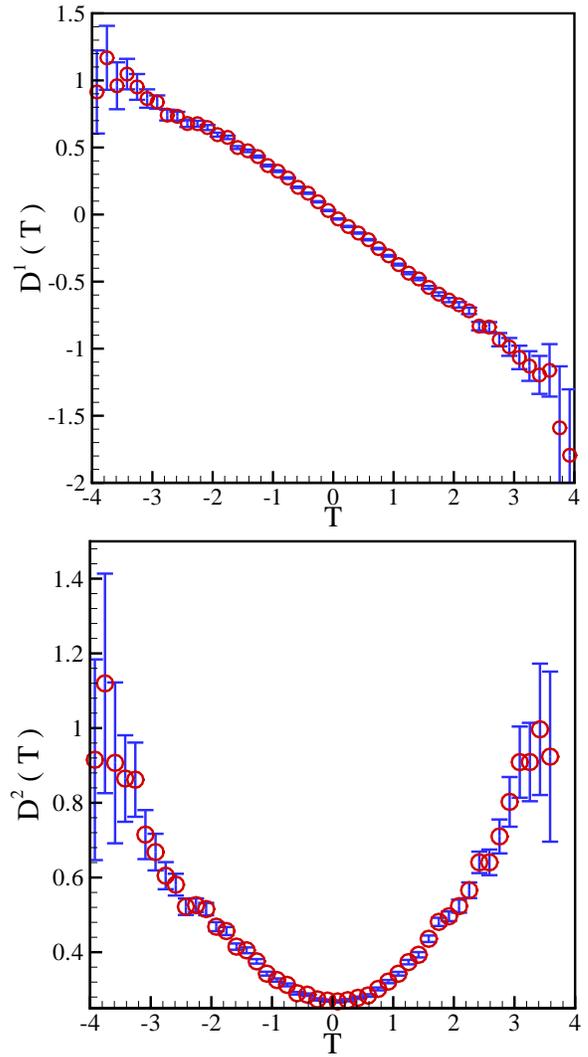}
 \narrowtext \caption{The drift and diffusion coefficients, $D^{(1)}(T)$ and $D^{(2)}(T)$
in Eq. (\ref{d12}), following, respectively, linear and quadratic
behaviors.} \label{f4}
\end{figure}

According to the Pawula's theorem \cite{ris84}, for a process with
$D^{(4)}\sim 0$, all the $D^{(n)}$ with $n\geq 3$ vanish, and the
KM expansion reduces to the FP equation, known also as the
Kolomogrov equation \cite{ris84}
\begin{equation}\label{fokker}
\frac{\partial}{\partial\phi}P(T,\phi)=\left[-\frac{\partial}{\partial
T}D^{(1)}(T,\phi)+ \frac{\partial^2}{\partial
T^2}D^{(2)}(T,\phi)\right]P(T,\phi)\;.
\end{equation}
Here $D^{(1)}(T,\phi)$ is the 'drift' term, describing the
deterministic part of the process, while $D^{(2)}(T,\phi)$ is the
`diffusion' term. Using the Ito interpretation, the FP equation is
equivalent to the following Langevin equation \cite{ris84}
\begin{equation}
\frac{d}{d \phi}T(\phi)=D^{(1)}( T ) + \sqrt{D^{(2)}( T )} \hskip
.1 cm f(\phi). \label{gen}
\end{equation}
Here, $f(\phi)$ is a $\delta$-correlated Gaussian random force
with zero mean (i.e., $<f(\phi) f(\phi')>=3D \delta(\phi-\phi')$).
For the WMAP data, the drift and diffusion coefficients $ D^{(1)}$
and $D^{(2)}$ are shown in Fig. \ref{f4}. It turns out that the
drift coefficient $D^{(1)}$ is a linear function in $T$, whereas
the diffusion coefficient $D^{(2)}$ is a quadratic function. For
large values of $T$, our estimates become poor and, thus, the
uncertainty increases. From the analysis of the data set, we
obtain the following approximate relations,
\begin{eqnarray}
\label{d}
&& D^{(1)} (T) =-0.330 T,  \cr \nonumber \\
&& D^{(2)}(T) =0.060 T^2+ 0.002 T + 0.270,
\end{eqnarray}
where we used the isotropy assumption which implies that $D^{(n)}
(T,\phi) =D^{(n)} ( T )$. The temperature field is measured in
units of its standard deviation. The fourth-order coefficient
$D^{(4)}$ is, in our analysis, $ {D^{(4)}} \simeq 10^{-2}
{D^{(2)}}$, so that we can ignore the coefficients $D^{(n)}$ for
$n \geq 3$. Furthermore, using Eq. (\ref{gen}), it becomes clear
that we are able to separate the deterministic and the noisy
components of the CMB fluctuations in terms of the coefficients
$D^{(1)}$ and $D^{(2)}$.

\section{The (non)-Gaussianity test of CMB}
\label{gaussian} Cosmological models of the structure formation
are based on the standard inflationary paradigm, which implies
Gaussian fluctuation of the density field
\cite{mu81,haw82,guth85,bar86}. The Gaussianity in the density
perturbation directly translates into the Gaussianity of the CMB
temperature fluctuations. However, along with the standard
inflationary models, there exist theories that predict the
non-Gaussianity of the primordial fluctuations. Inflation with two
or more scalar fields can provide significant deviations from the
Gaussianity \cite{linde97,pee99a,pee99b,anto97}. Another
possibility is the manipulation of the CMB data after the
recombination, due to subsequent weak gravitational lensing
\cite{fuk95,ber97} and various foregrounds, such as dust emission,
synchrotron radiation, or unresolved point sources \cite{ban96}.
One should also take into account the additional instrumental
noise in the observational data \cite{teg97}.

There are standard methods of searching for non-Gaussian signature
in the CMB data, such as using peak distributions
\cite{bar86,bond87}, the genus curve \cite{col88,smoot94}, peak
correlations \cite{kog96} and global Minkowski functional methods
\cite{win97}. More recent methods, such as a technique for studying
hydrodynamic turbulence and detecting non-Gaussianity, and fractal
analysis, can also be used for the CMB data \cite{sree03,sadegh06}.
Moreover, the Gaussianity of the CMB at different angular scales
have been tested
\cite{heavens98,schmalzing98,ferreira98,pando98,bromley99,banday00,contaldi00,mukherjee00,magueijo00,novikov00,sandvik01,magueijo01,barreiro00,phillips01,komatsu02,komatsu01,kunz01,aghanim01,cayon02,park01,shandarin02,wu01,santos02,polenta02}.
Most of the previous works were based on the consistency between the
CMB data and simulated Gaussian realizations. So far, they have
found no significant evidence for cosmological non-Gaussianity.

We note that for a Gaussian distribution, all the even moments are
related to the second one through $\langle
T^{2n}\rangle=\frac{2n!}{2^nn!}\langle T^2\rangle^n$ (e.g., for
$n=2$, $\langle T^{4}\rangle=3\langle T^2\rangle^2$), while the odd
moments are zero. We checked directly the relation between the
moments for the CMB data. The results are summarized in Table
\ref{Tb1}. The ratios $\langle T^4\rangle/(3\langle T^2\rangle^2)$
are, $1.077\pm 0.004$ , $1.085\pm0.005$ and $1.048\pm0.005$, for the
entire data set, and the northern and southern hemispheres,
respectively. The odd moments $\langle T^3\rangle$ and $\langle T^5
\rangle$ are not zero for different parts of the CMB data (see Table
\ref{Tb1}). The odd moments of the temperature fluctuations can also
be used as the indicators for the hot and cold spots. The southern
part has a negative third moment, indicating greater prevalence of
the cold spots, whereas the northern hemisphere has a positive third
moment, indicating domination of the hot spots. The ratio of the
higher moments, such as $ \langle T^6\rangle/(15 \langle
T^2\rangle^3)$, are, $1.398\pm 0.026$, $1.396\pm 0.022$ and
$1.313\pm 0.035$ (for a Gaussian distribution this ration is unity)
for the entire data set, and the northern and southern hemispheres,
respectively.

The ratio of the higher moments, such as $\langle T^6\rangle/(15
\langle T^2\rangle^3)$, for the entire sky, and the northern and
southern hemispheres are larger than those for a Gaussian
distribution, implying that the probability distribution function
for the temperature fluctuations in the CMB has fat tails compared
to a Gaussian distribution.

\begin{figure}[t]
\vspace{-1cm}
 \epsfxsize=9truecm\epsfbox{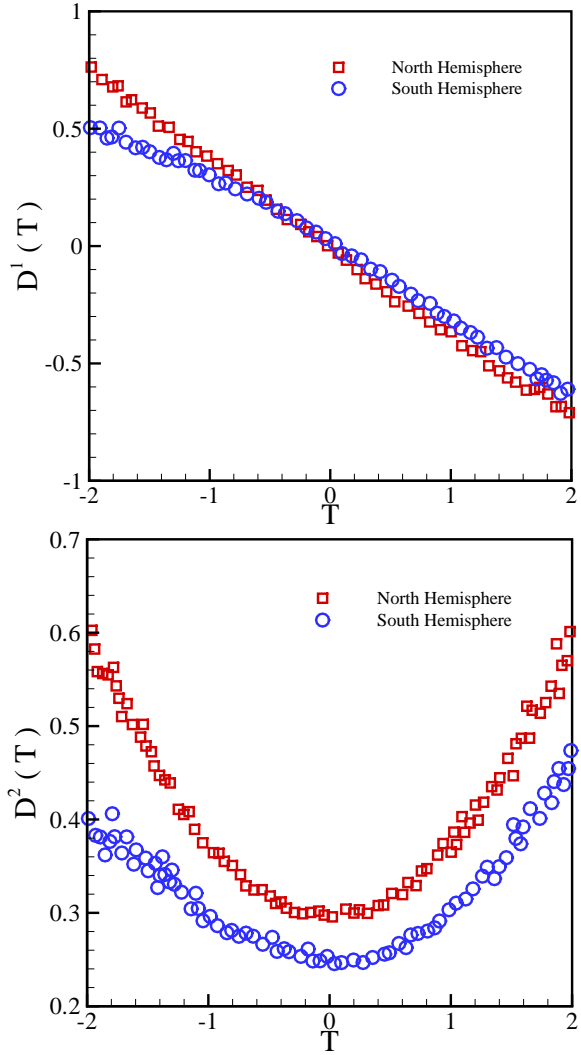}
 \narrowtext \caption{ Drift and diffusion coefficients $D^{(1)}(T)$ and
$D^{(2)}(T)$ for the northern and the southern hemisphere parts of
the CMB data.} \label{f5}
 \end{figure}

Let us examine the predictions for the moments of the temperature
fluctuations via the FP equation, and compare their values with
the direct evaluation of results presented in Table \ref{Tb1}.
Using the stochastic properties of the temperature field, one can
check the deviation of the fluctuations from the Gaussian
distribution in the northern and southern hemispheres, as well as
in the entire data. As pointed out in section \ref{M_P}, the PDF
of the temperature fluctuations satisfies the KM expansion. Using
Eq. (\ref{fokker1}) we calculate the $nth$ moment of the
temperature fluctuations by multiplying both sides of Eq.
(\ref{fokker1}) by $T^n$ and integrating over the temperature:
\begin{equation}
\label{moment} \frac{d}{d\phi}\langle
T^n\rangle=\sum_{m=1}^{\infty} \frac{n!}{(n-m)!}\langle
D^{(m)}(T)T^{n-m}\rangle.
\end{equation}
We set $n=4$ in the above equation and find the equation for the
fourth moment to be
\begin{eqnarray}
\label{moment4} &&\frac{d}{d\phi}\langle T^4\rangle=4\langle
D^{(1)}(T)T^3\rangle+12 \langle
D^{(2)}(T)T^2\rangle\nonumber\\
&&\qquad +24\langle D^{(3)}(T)T\rangle+24\langle
D^{(4)}(T)\rangle.
\end{eqnarray}
The KM coefficients for the northern and the southern hemispheres
are given in Table \ref{Tb2}. For the isotropic case, all the
moments of temperature fluctuations are independent of $\phi$, and
the left-hand side of Eq. (\ref{moment4}) vanishes. Using the
results presented in Table \ref{Tb2}, for the northern and the
southern hemispheres, Eq. (\ref{moment4}) reduces to
\begin{eqnarray}
\label{ns}\langle T^4\rangle_{north}&=&(2.71\pm 0.06)\langle
T^2\rangle^2+(0.028\pm 0.010)\langle T^3\rangle \sqrt{\langle
T^2\rangle},\nonumber \\ \langle T^4\rangle_{south}&=&(3.27\pm
0.09)\langle T^2\rangle^2-(0.017\pm 0.020)\langle T^3\rangle
\sqrt{\langle
T^2\rangle},\nonumber\\
\end{eqnarray}
where for the northern and southern hemispheres, $\langle
T^3\rangle=(1.283\pm 0.125)\times 10^{-5}$, $\sqrt{\langle
T^2\rangle}=(7.010\pm 0.004)\times10^{-2}$ and $\langle
T^3\rangle=(-5.408\pm 0.175)\times 10^{-5}$, $\sqrt{\langle
T^2\rangle}=(7.914\pm 0.004)\times10^{-2}$, respectively. We
repeated the same procedure for the entire data set and obtained
the following relation between the moments
\begin{eqnarray}
\label{whole}\langle T^4\rangle=(3.07\pm0.07)\langle
T^2\rangle^2+(0.034\pm 0.020)\langle T^3\rangle \sqrt{\langle
T^2\rangle},
\end{eqnarray}
where $\langle T^3\rangle=(-2.032\pm 0.108)\times 10^{-5}$ and
$\sqrt{\langle T^2\rangle}=(7.475\pm 0.003)\times 10^{-2}$. The
drift and diffusion coefficients of the northern and the southern
hemisphere are shown in Fig.\ \ref{f5}.

As mentioned earlier, for a Gaussian distribution, all the even
moments are related to the second one, while the odd moments are
zero. Eqs. (\ref{ns}) and (\ref{whole}) show that we have
deviations from the Gaussianity, resulting from the coefficient of
$\langle T^2 \rangle^2$. 
Note that the result presented by Eq. (\ref{whole}) are compatible
with direct calculation of moments for different part of the CMB
data.

\section{Markov Nature of CMB and Angular Power Spectrum}
\label{KMC} Using statistical isotropy, the two-point temperature
correlation function is expanded in term of the Legendre functions
as
\begin{eqnarray}
\langle T (\hat n_1) T (\hat n_2) \rangle=\sum _l \frac{2l+1}{4
\pi}C_l P_l (\cos(\Theta)),
\end{eqnarray}
where $C_l$ is the angular power spectrum, $P_l$ is the Legendre
polynomial of order $l$ and $\Theta=\arccos(\hat n_1.\hat n_2)$.
To find the relation between the Markovian properties of the CMB
data and the $C_l$'s, we start with the temperature-increment
moments, $S_2 = \langle [ T(\hat n_1)- T(\hat n_2) ]^2 \rangle$.
The second moment $S_2$ allows us to determine $\langle T(\hat
n_1) T(\hat n_2) \rangle$ and, hence, the $C_l$'s. On the other
hand, $S_2$ can be obtained through $P( \Delta T, \Theta )$, where
$\Delta T =T(\hat n_1) - T(\hat n_2)$. The FP equation
\cite{ris84} for $\Delta T(\Theta) $ is given by,
\begin{eqnarray}
\label{delp}
&&\frac {d}{d \Theta} P(\Delta T,\Theta)=\nonumber\\
&&[-\frac{\partial }{\partial \Delta T} D^{(1)}(\Delta T, \Theta)
+\frac{\partial^2 }{\partial \Delta T^2} D^{(2)}(\Delta T,
\Theta)] P(\Delta T, \Theta).
\end{eqnarray}
From the analysis of the CMB data we obtain the following
equations for $D^{(1)}(\Delta T, \Theta)$ and $D^{(2)}(\Delta T,
\Theta)$
\begin{eqnarray}\label{18}
&& D^{(1)}(\Delta T, \Theta)=(-0.190-\frac{0.182}{\Theta})
\hskip .2 cm \Delta T,  \cr \nonumber \\
&&D^{(2)}(\Delta T,\Theta)=
[0.021+0.025 \exp({-\frac{\Theta}{8.896}})] (\Delta T)^2\nonumber\\
&&\hspace{2.2cm}+ 0.279+\frac{0.014}{\Theta^{0.429}}.
\end{eqnarray}
Here, $\Delta T$ is measured in units of the standard deviation of
$T$. Equation (\ref{delp}) allows us to obtain an equation for the
second moment of the temperature increments, $S_2$. Multiplying
Eq. (\ref{delp}) by $\Delta T ^2$ and integrating over the
increment of the temperature yield
\begin{eqnarray}
\label{para} \frac{d}{d\Theta}\langle
&&T(\hat{n_1})T(\hat{n_2})\rangle=2[\alpha(\Theta)+\omega(\Theta)]\langle
T(\hat{n_1})T(\hat{n_2})\rangle\nonumber\\&&\qquad-\sigma^2[2\alpha
(\Theta)+2\omega(\Theta)-\beta(\Theta)],
\end{eqnarray}
where $\sigma^2=\langle T^2\rangle$, $D^{(1)}(\Delta T,
\Theta)=\alpha(\Theta) \Delta T(\Theta)$, and $D^{(2)}(\Delta T,
\Theta)=\beta(\Theta)\sigma^2+\lambda(\Theta)\sigma\Delta
T(\Theta)+\omega(\Theta)\Delta T(\Theta)^2$. The coefficients
$\alpha(\Theta)$, $\beta(\Theta)$, $\lambda(\Theta)$, and
$\omega(\Theta)$ are given by Eq. (\ref{18}).

\begin{figure}
\vspace{-1cm}\epsfxsize=9truecm\epsfbox{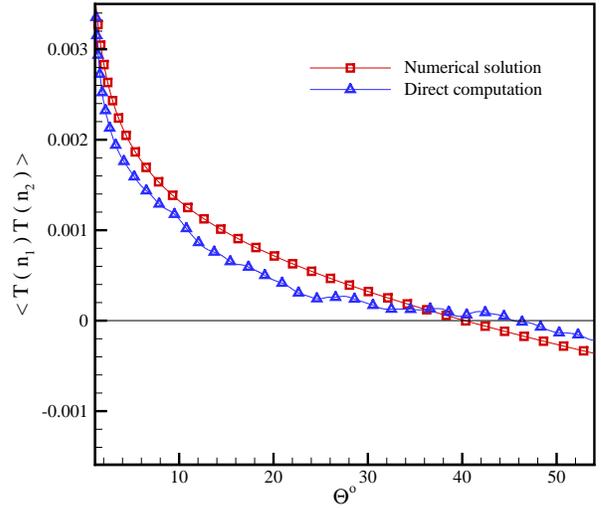} \narrowtext
\caption{ Comparison of the numerical solution of Eq. (\ref{para})
and direct computation of $\langle
T(\hat{n_1})T(\hat{n_2})\rangle$.} \label{f6}
 \end{figure}

We obtain the correlation functions through numerical solution of
Eq. (\ref{para}) and compare it with the direct correlation shown
in Fig. \ref{f6}. The two approaches yield similar behavior. At $
\Theta \simeq 39^o$ the correlation function vanishes. The
relation between the KM coefficients with $C_{l}$ is then obtained
by expanding the solution
of Eq. (\ref{para}) in terms of $P_{l}$.\\

\section{Conclusion}
\label{conc} We studied the stochastic nature of the temperature
fluctuations in the CMB. The Markov angle, as the characteristic
scale of the Markov properties of the CMB data, was obtained.
According to the theory of stochastic process, the CMB data at
scales larger than the Markov angle can be considered as a Markov
process. This means that the data located at separations larger
than the Markov scale can be described as a Markov chain. We also
obtained the angular scale correlation for the CMB data and showed
that it is independent of the Markov angular scale. This point can
be explained as a simple example of the Brownian motion. Its
dynamics is described by the Langevin equation, $\frac{dv}{dt}
=-\alpha v + \eta(t)$ (similar to Eq. (\ref{gen})), where the
force $\eta(t)$ is a zero-mean Gaussian white noise. It is known
that this process is a Markov process with a Markov time scale of
unity \cite{ris84}. On the other hand, the correlation time scale
is of the order of $1/\alpha$, which is the characteristic time
scale in the Langevin equation \cite{ris84}. This means that for a
Markovian process the Markov time scale does not depend on the
correlation time scale.

We showed that the probability density of the temperature
increments satisfies a Fokker-Planck equation that encodes the
Markov property of the fluctuations. We gave the expressions for
the Kramers-Moyal coefficients of the stochastic process $T$ and
$\Delta T(\Theta)$, using the polynomial anzats
\cite{fri97,fri00,fri97b,dav99,rag01,fri02}. The Fokker-Planck
equation enables us to derive a simple equation that governs the
phenomenon in terms of azimuthal spherical coordinate. One of the
important points that can be tested by this method is the
(non)-Gaussianity of the temperature field.

Using the Fokker-Planck equation, we obtained the relation between
the fourth and the lower moments, and observed small deviations
from the Gaussianity. The same calculation was carried out by
dividing the data into the northern and southern hemispheres. The
PDF of the temperature exhibits fat tails for the northern and the
southern hemispheres. The third moment indicates that, we have hot
spots in the north, in contrast with the cold spots in the south
hemisphere. To show a link between our method and the standard
analysis of the CMB data through $C_{l}$ calculation, we obtained
the evolution equation for the correlation function through the
Fokker-Planck equation that governs the temperature increments. We
found good agreement between our method and that of direct
correlation function calculation.

The most important result of this paper is a possible
interpretation of the CMB data in terms of the Markov angular
scale. In this work we computed the Markov angular scale for the
temperature fluctuations in the CMB, in the range of
$1.01^{+0.09}_{-0.07}$, where comparison with the event horizon
shows that the two angles are of the same order of magnitude. A
possible interpretation may be the physical connection of the
Markov angular scale for the CMB data to the event horizon at the
last scattering surface.

From the definition of the event horizon, two points located
further apart than the Markov angular scale do not connect
gravitationally. From the inflationary scenario for the early
universe, all the points in the universe have been correlated
during the inflationary epoch, and get almost uniform
Harrison-Zel'dovich spectrum. After the end of the inflation, the
perturbation remained from the inflation started to grow and the
gravitational effect of each point could travel within the event
horizon scales, so that outside the horizon we have the primordial
spectrum from the inflation, while inside the horizon this
spectrum has been vanished due to the gravitational interaction
between the points. The result is that, if we look at the density
perturbation in the CMB as a stochastic field, the conditional
probability of finding two points inside the event horizon scale
will depend on all the particles in between, while for the points
outside the event horizon, the memory of scale-invariant spectrum
should be restored. A better interpretation of this issue needs a
$N$-body simulation of structure formation from the inflationary
epoch to the last scattering surface, together with a simultaneous
calculation of the Markov length.

We would like to acknowledge the WMAP team for providing us with
their data. We thank R. Ansari, H. Arfaei, V. Karimipour, R.
Mansouri, N. Taghavinia, M. A. Vesaghi and M. Sahimi for useful
discussions. This work was supported by the Research Institute for
Astronomy and Astrophysics of Maragha-Iran.

\newpage
\begin{widetext}
\begin{table}
\caption{\label{Tb1}The values of Moments $\langle T^n \rangle$ for
whole sphere, northern and the southern hemisphere of the CMB map
data.}
\medskip
\begin{tabular}{|c|c|c|c|c|}
  & $\langle T^2 \rangle$ & $ \langle T^3 \rangle$ & $ \langle T^4\rangle $ &$ \langle T^5\rangle $  \\
  \hline
{\rm Whole-sphere} & $(5.588 \pm 0.005) \times 10^{-3}$& $(-2.033
\pm 0.108 )\times 10^{-5}$ & $(1.009 \pm 0.003)\times 10^{-4}$ &
$(-1.298\pm
0.132)\times 10^{-6}$ \\
\hline
  {\rm Northern-hemisphere} & $(4.914 \pm 0.006) \times 10^{-3}$&
  $(1.283\pm 0.125)\times 10^{-5}$ & $(7.867 \pm 0.0305)\times 10^{-5}$ & $(7.632\pm
1.028)\times 10^{-7}$ \\
  \hline
 {\rm Southern-hemisphere}  & $(6.264  \pm 0.007)\times 10^{-3}$& $(-5.408 \pm
0.175 )\times 10^{-5}$ & $(1.233 \pm 0.005)\times10^{-4}$ &
$(-3.392\pm 0.244)\times 10^{-6}$
 \\
\end{tabular}
\end{table}
\end{widetext}

\begin{widetext}
\begin{table}
\caption{\label{Tb2}The values of Kramers-Moyal coefficients for the
northern and the southern hemisphere of the CMB map data.}
\medskip
\begin{tabular}{|c|c|c|c|c|}
  & $D^{(1)}(T)$ & $D^{(2)}(T)$ & $D^{(3)}(T)$ &$ D^{(4)}(T)$ \\
  \hline
  {\rm Northern-hemisphere} & -0.370$T$ & 0.290-0.003$T$+0.075$T^2$ & -0.001-0.110$T$
&0.040-0.007$T$+0.0182$T^2$ \\
   & & &+0.003$T^2$-0.010$T^3$ & -0.0005$T^3$+0.001$T^4$
\\\hline
 {\rm Southern-hemisphere} & -0.290$T$ & 0.250+0.007$T$+0.047$T^2$ & 0.004-0.078$T$ &
0.031-0.003$T$+0.020$T^2$ \\
   & &  & -0.005$T^2$-0.003$T^3$ & +0.001$T^3$-0.001$T^4$ \\
\end{tabular}
\end{table}
\end{widetext}

\end{document}